\documentclass[aps,prd,twocolumn,showpacs,floatfix,preprintnumbers,amsfont,amsmath,amssymb,nofootinbib, superscriptaddress]{revtex4-1}

\usepackage{color}
\usepackage{hyperref}
\usepackage[normalem]{ulem}
\usepackage{lipsum}
\usepackage{url}
\usepackage{float}
\usepackage{ctable}
\usepackage{graphicx}
\usepackage[utf8]{inputenc}
\usepackage[font=small, justification=raggedright, format=plain
  ]{caption} 
\usepackage{subcaption}
\usepackage{mathtools}
\usepackage{bm}
\usepackage{siunitx}

\newcounter{magicrownumbers}
\newcommand\rownumber{\stepcounter{magicrownumbers}\arabic{magicrownumbers}}

\newcommand{\sk}[1]{}

\newcommand{\reffig}[1]{Fig.~\ref{fig:#1}}

\newcommand{\snr}{{\rm SNR}}
\newcommand{\rhoh}{\rho_{\rm H}}
\newcommand{\rhol}{\rho_{\rm L}}

\def\Thintr{\mathbf{\Theta_{\rm intr}}}
\def\Thext{\mathbf{\Theta_{\rm ext}}}
\def\bfd{\boldsymbol{d}}

\def\bfd{\boldsymbol{d}}

\newcommand{\be}{\begin{equation}}
\newcommand{\ee}{\end{equation}}
\newcommand{\ba}{\begin{eqnarray}}
\newcommand{\ea}{\end{eqnarray}}

\newcommand{\apjs}{Astrophys. J. Suppl. Ser.}

\newcommand{\mnras}{Mon. Not. R. Astron. Soc.}

\allowdisplaybreaks

\begin{document}

\title{Detecting Gravitational Waves With Disparate Detector Responses: \\ Two New Binary Black Hole Mergers}
\author{Barak Zackay}
\email{bzackay@ias.edu}
\affiliation{\mbox{School of Natural Sciences, Institute for Advanced Study, 1 Einstein Drive, Princeton, NJ 08540, USA}}

\author{Liang Dai}
\affiliation{\mbox{School of Natural Sciences, Institute for Advanced Study, 1 Einstein Drive, Princeton, NJ 08540, USA}}
\author{Tejaswi Venumadhav}
\affiliation{\mbox{School of Natural Sciences, Institute for Advanced Study, 1 Einstein Drive, Princeton, NJ 08540, USA}}
\author{Javier Roulet}
\affiliation{\mbox{Department of Physics, Princeton University, Princeton, NJ, 08540, USA}}

\author{Matias Zaldarriaga}
\affiliation{\mbox{School of Natural Sciences, Institute for Advanced Study, 1 Einstein Drive, Princeton, NJ 08540, USA}}

\date{\today}


\begin{abstract}

We introduce a new technique to search for gravitational wave events from compact binary mergers that produce a clear signal only in a single gravitational wave detector, and marginal signals in other detectors.
\sk{with a clear signal in one detector and marginal signals in others. Such a situation can arise due to sensitivity differences between different detectors or due to unfavorable source sky position and orientation. }
Such a situation can arise when the detectors in a network have different sensitivities, or when sources have unfavorable sky locations or orientations.
We start with a short list of loud single-detector triggers from regions of parameter space that are empirically unaffected by glitches (after applying signal-quality vetoes).
\sk{from ``glitch-free" regions of the parameter space,}
For each of these triggers, we compute evidence for astrophysical origin from the rest of the detector network by coherently combining the likelihoods from all detectors and marginalizing over extrinsic geometric parameters.
\sk{It implements a fully coherent fitting procedure applied to the LIGO/Virgo detector network using Bayesian marginalization over the extrinsic parameters applied to a short list of loud candidates in a ``glitch-free" part of the template bank. }
We report the discovery of two new binary black hole (BBH) mergers in the second observing run of Advanced LIGO and Virgo (O2), in addition to the ones that were reported in \cite{LIGOScientific:2018mvr} and \cite{Venumadhav:2019lyq}. We estimate that the two events have false alarm rates of one in 19 years (60 O2) and one in 11 years (36 O2).

\sk{We found a simple, general criterion which given a list of significant single detector triggers outputs a short list on which the top candidates are confirmed as genuine astrophysical events based on counterpart signals in other detectors. These events were found in a refined analysis focusing on the case in which the two LIGO detectors have very different signal responses. The new analysis recovers several events that were previously found as coincident triggers, and additionally yields two new events with false alarm rates of one in 60 O2's and one in 36 O2's.}

One of the events, GW170817A, has primary and secondary masses $m_1^{\rm src} = 56_{-10}^{+16} \, M_\odot$ and $m_2^{\rm src} = 40_{-11}^{+10} \, M_\odot$ in the source frame. 
The existence of GW170817A should be very informative about the theoretically predicted upper mass gap for stellar mass black holes.
Its effective spin parameter is measured to be $\chi_{\rm eff} = 0.5 \pm 0.2$, which is consistent with the tendency of the heavier detected BBH systems to have large and positive effective spin parameters.
The other event, GWC170402, will be discussed thoroughly in future work.

\sk{a binary total mass of $M^{\rm src}_{\rm tot} = 98^{+17}_{-11}\,M_{\odot}$ which would make it the most massive binary merger observed so far. Nitz's candidate is now out in the wild, so can we say that? Is this really heavier than 170729? puts to test theories about a possible upper mass gap for stellar mass black holes.}

\end{abstract}

\maketitle

\section{Introduction}

The LIGO-Virgo Collaboration (LVC) detected ten binary black hole (BBH) coalescence events during their first and second observing runs (O1 and O2)~\cite{LIGOScientific:2018mvr}. 
We performed an independent analysis of the publicly released O1 and O2 data, and reported seven additional BBH events in Refs.~\cite{pipeline, Venumadhav:2019lyq, GW151216}. 
Several of the events we identified were recently also found in an independent search using the PyCBC analysis pipeline, which also reported a new massive BBH~\cite{Nitz:2019hdf}. 

\begin{figure}[!h]
    \centering
    \includegraphics[width=\linewidth]{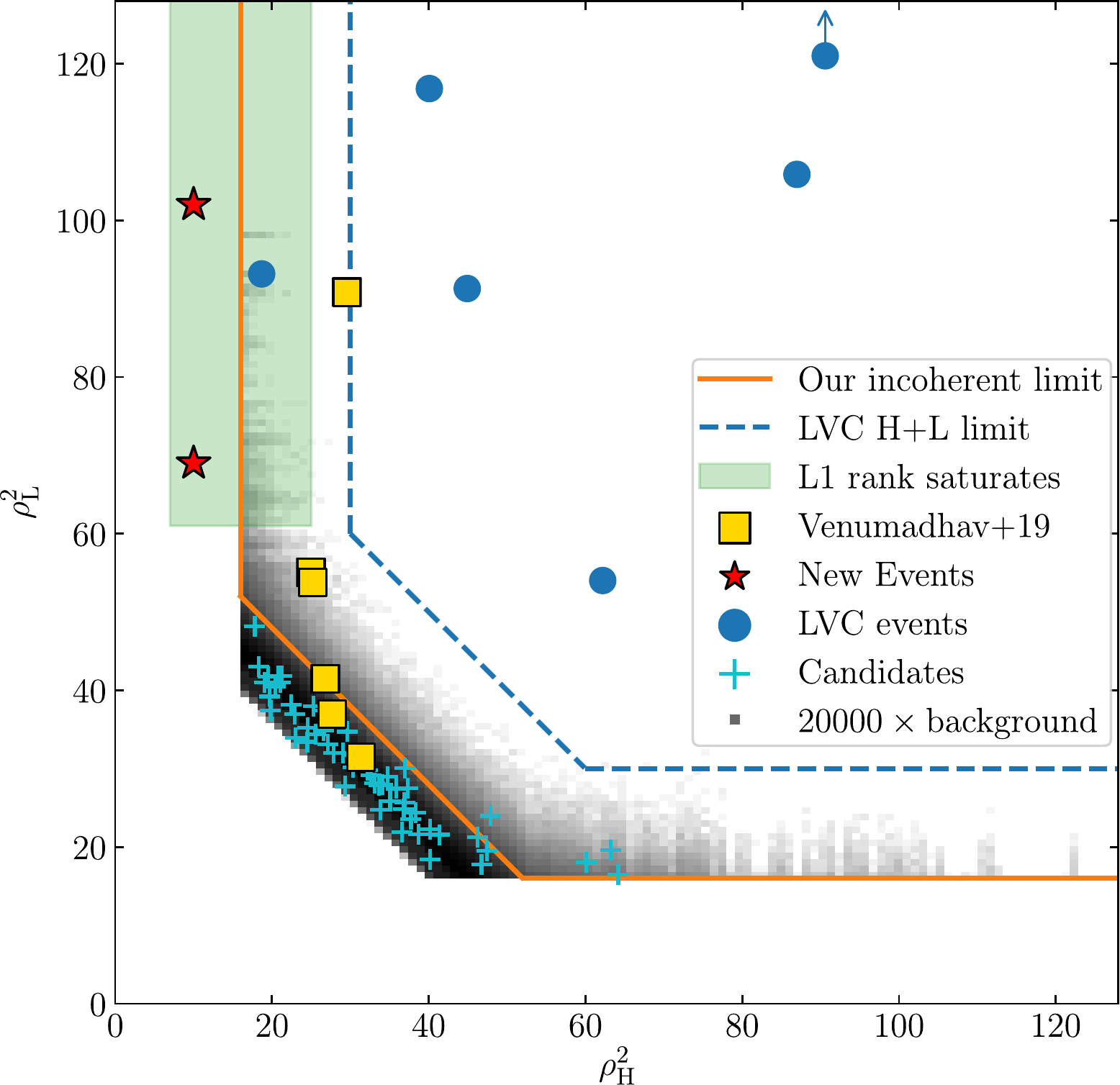}
    \caption{\label{fig:money-plot34} Incoherent Hanford (H1) and Livingston (L1) $\snr^2$ for coincident and background triggers (computed using \num{20000} time slides), for all the sub-banks with events. The blue and orange lines are approximate incoherent detection limits for analyses in Refs.~\cite{LIGOScientific:2018mvr} and \cite{Venumadhav:2019lyq}, respectively, restricted to using H1 and L1 data only. \sk{The incoherent detection limits are inexact, since they do not account for the consistency of the time-delays, phases, and the relative sensitivities of the detectors. We show them only to qualitatively assess the relative volumes probed by different analyses.} GW170814 has $\rho_{\rm L}^2=170$, higher than shown here (indicated with an arrow), and GW170608 is not shown because its H1 data is not part of the bulk O2 data release. Figure adapted from Ref.~\cite{Venumadhav:2019lyq}. }
\end{figure}

Figure \ref{fig:money-plot34} summarizes the sensitivity reach of both search efforts, in terms of the signal-to-noise ratios (SNR) in the Hanford (H1) and Livingston (L1) detectors. 
In this paper, we extend our search to cover the region in parameter space in which the signal response is very high in one detector, but small in the other (this regime is shown as the teal region\footnote{The analogous region with high SNR in H1 corresponds to much smaller sensitive volume} in Fig.~\ref{fig:money-plot34}. 
It is in general challenging to reliably compute the false alarm rate (FAR) of a trigger in this region, because throughout the entire observing run, there are only a small number of triggers that (a) have comparably high SNRs, (b) are well fit by similar waveforms, and (c) pass our vetoes. 
The fact that we cannot realistically simulate interferometer data prevents us from empirically measuring the FAR.

We could empirically measure the number of fainter triggers and extrapolate the distribution.
However, loud triggers are mainly produced due to anomalous detector behavior, i.e., so-called glitches~\cite{blackburn2008lsc, abbott2016characterization}, rather than stationary Gaussian random noise. 
Since we do not completely understand the physical or instrumental origins of glitches~\cite{Cabero:2019orq}, extrapolations of their distributions to higher values of SNR come with significant uncertainty.

In our previous analysis in Ref.~\cite{Venumadhav:2019lyq}, we did not search for events with highly incommensurate SNRs in the two LIGO detectors (the region of phase space with $\rhol^2 > 66$ and $\rhoh^2<16$, where $\rho$ is the SNR and the subscript refers to the detector), since we imposed a threshold on the SNR for collecting triggers (this threshold equaled 4 in the banks covering massive BBH mergers). 
Given the higher sensitivity of L1 compared to H1 on average throughout O2, we estimated that our cut on single-detector SNR reduced the sensitive volume of our search by $\sim 10\%$. 
Note that even when the two LIGO detectors have comparable sensitivity, astrophysical sources with unfortunate sky positions and orbital orientations can produce signals of different strengths in the detectors. 
The search described in this paper revealed two additional interesting BBH triggers that are above the thresholds of significance for being called events~\cite{LIGOScientific:2018mvr}.

One event, GW170817A (not to be confused with the binary neutron star merger event GW170817~\cite{GW170817}) comes from a pair of black holes with a very high total mass $\sim 100\,M_\odot$. 
The existence of such a BBH system is informative about theories of the evolution and death of massive stars, which generically predict an upper mass gap at $\sim 50\,M_\odot$ for stellar mass black holes.

Efforts to estimate the parameters of the other candidate, GWC170402, yield significant evidence that the signal is not fully described by waveforms of the dominant harmonic mode for circular binaries with aligned spins. 
This will be discussed thoroughly in a forthcoming paper. 
At the time of writing, we do not have a waveform model that completely models the observed signal, and hence we do not attach a ``GW'' prefix but a temporary GWC for ``GW candidate''.

The paper is organized in the following way:
In Section \ref{sec:Methodology} we discuss the technique and its application to the O2 data. 
In Section \ref{sec:results} we summarize the analysis results.
In Section \ref{sec:astro} we present the results of parameter estimation for GW170817A, and discuss its relevance to astrophysical formation scenarios for massive black holes.
We finish with our conclusions in Section \ref{sec:conclusions}, and provide some extra details in the appendices.

\section{Methodology and Results} 
\label{sec:Methodology}

In this section, we describe the methods we use to analyze and assign significance to single detector triggers, and present results alongside.
We begin with a brief overview below, and expand upon the details in individual sections.

In this analysis, we focus on triggers from template banks for relatively massive BBH mergers, with detector-frame chirp masses $\geq 20\,M_{\odot}$.
This part of parameter space is most promising for the search presented in this paper, because it contains twelve of the BBH mergers that have been detected as coincident H1 and L1 triggers so far, and hence there is a significant chance that one or more events from similar sources may have been missed by previous analyses (due to SNR cuts we imposed when collecting triggers, or approximations used for the coherent score that were valid in the high SNR regime).

To perform this specialized search, we first define a set of significant L1 triggers, which are so loud that it is extremely unlikely that Gaussian random noise produces them, even over the entire length of the O2 run.
However, glitches can produce such loud triggers, and hence we rank these triggers not by their SNR, but by an empirical measure of how frequently known glitches contaminate their surrounding phase space.
Section \ref{sec:GlitchSimilarityScore} presents details of this ranking and justification for it. 

We then examine the strain data from the less-sensitive detectors (H1 and/or V1 (i.e., Virgo)) for counterpart signals of each of the above L1 triggers.
We define a score that, given a signal in L1, coherently computes the joint likelihood from the data in all available detectors, and marginalizes over extrinsic parameters of the source. 
We derive this score and its properties in Section \ref{sec:FishingScore}.

We next need to combine the information from the more- and less-sensitive detectors (L1, and H1 and/or V1, respectively) and estimate a final false alarm rate (FAR) for the triggers. We describe our method to do so in Section \ref{sec:FAR}.

The FAR quantifies the rate at which detector noise produces triggers above a threshold. In a similar manner to searches of coincident triggers, we need to compare this rate to the rate at which the known astrophysical population of mergers would produce the triggers, and estimate the probability of astrophysical origin ($p_{\rm astro}$) for the candidates. Section \ref{sec:PASTRO} outlines our method to accomplish this.

Finally, in Section \ref{sec:170818}, we validate our methods by applying them to the event GW170818, which is a highly significant GW event in the official catalog released by the LVC \cite{LIGOScientific:2018mvr} that lies in the region of phase-space covered by this search (the blue circle within the teal region in Fig.~\ref{fig:money-plot34}).

\subsection{Ranking Single Detector Triggers} \label{sec:GlitchSimilarityScore}

\begin{table*}[!ht]
    \centering
    \begin{tabular}{|c|c|c|c|c|c|c|c|c|}
    \hline
        \rule{0pt}{3ex} L1 rank &GPS time & $\rhol^2$ & \# similar triggers   & $C(\mathcal{S} | H_0)$ &  $C(\mathcal{S} | H_1)$ & $\frac{P(\mathcal{S} | H_1)}{P(\mathcal{S} | H_0)}$ &Comment\\[1ex]
        \hline
        \rownumber &1187058327.068& 93.1& 0 & $<10^{-3}$ & 0.16 & 37 & GW170818\footnote{For the purpose of demonstrating our new methodology, we present numbers corresponding to analyzing data from only the two LIGO detectors, even though Virgo detected GW170818 at ${\rm SNR} \simeq 4$~\cite{LIGOScientific:2018mvr}.}\\
        \rownumber &1187529256.504& 92.1& 0 & - & - & -  &GW170823\\
        \rownumber &1169069154.564& 90.8& 0 & - & - & -  &GW170121\\
        \bf \rownumber  &{\bf 1175205128.565}& {\bf 72.9}& {\bf 0} & {\bf 0.015} & {\bf 0.022} & {\bf 0.547} & {\bf GWC170402}\\
        \rownumber &1186741861.51& 174.6& 1 & - & - & -  &GW170814\\
        \rownumber &1167559936.584& 107.3& 1 & - & - & - &GW170104\\
        \rownumber &1186302519.731& 118.6& 2 & - & - & - &GW170809\\
         
         \bf \rownumber  &{\bf 1186974184.716}& {\bf 98.5}& {\bf 5} & {\bf 0.028} & {\bf 0.055} & {\bf 0.98} & {\bf GW170817A}\\
        \rownumber &1174043898.842& 75.7& 9 & 0.36 & 0.001  & 0.008 & Background \\
        \rownumber &1170885005.109& 66.4& 16 & 0.49 & 0.003 & 0.013 & Background \\
        \rownumber &1178083239.592& 74.4& 22 & 0.34 & 0.003 & 0.016 & Background \\
        Removed\footnote{We removed this candidate as its Livingston spectrogram shows immediately obvious signs of non-stationary activity, or `glitchy' behavior (see Fig.~\ref{fig:GlitchSpecGram} in Appendix~\ref{sec:glitch_specgram}). We include it in the list for completeness.} &1173477193.704& 69.2& 1 & 0.38 & 0.014 & 0.011 & Artifacts present\\

\hline
    \end{tabular}
    \caption{Triggers ranked solely based on data from the Livingston (L1) detector. 
    The ranking is based on the number of similar triggers with L1 $\snr^2 = \rhol^2 > 55$ that pass our vetoes, which assesses the relative tendency of glitches in L1 to produce similar spurious background triggers. 
    Note that this simple ranking marks essentially all previously confirmed loud ($\rhol^2>60$) BBH mergers based on the L1 triggers alone. 
    The next three columns quantify the evidence for the astrophysical nature of the triggers from data in the Hanford (H1) detector, in terms of our coherent score $\mathcal{S}$ (see Eq.~\eqref{eq:fishS}): $C(\mathcal{S}|H_0)$ ($C(\mathcal{S}|H_1)$) is the probability of obtaining a coherent score higher (lower) than that of the trigger in a random segment of H1 data without a signal (with an injected signal with consistent intrinsic parameters).
    Note that the new triggers (marked bold) that have high ranks based on L1, also have significantly low values of false alarm probability, $C(\mathcal{S}|H_0)$.}
    \label{tab:SimilarityRanking}
\end{table*}

The L1 detector was more sensitive over most of the O2 run, and hence we expect that loud single detector events in H1 (with $\rhol^2 < 16$) are much rarer than similar events in L1. 
Moreover, we empirically observe that the L1 detector produces a much lower number of loud triggers that pass our signal-quality vetoes (i.e., glitches). 
Hence, we focus our efforts toward characterizing loud L1 triggers.

Our previous search within coincident triggers used rank functions to sort triggers from the two LIGO detectors by their significance~\cite{Venumadhav:2019lyq}. 
Rank functions empirically quantify the probability that the underlying noise process produces triggers at a given value of SNR; we computed them separately for each detector, and for different regions of the source parameter space. 
In particular, our search used several template banks (logarithmically-spaced in chirp mass), each in turn divided into sub-banks that captured the variety of waveform amplitude profiles \cite{templatebankpaper}. 
We computed rank functions separately for each sub-bank, since the non-Gaussian tails of the single-detector trigger distribution varied significantly as a function of parameters. 
This allowed us to assess the significance of coincident triggers by consistently and locally estimating the effects of glitches.

In Ref.~\cite{Venumadhav:2019lyq}, we noted that the rank functions empirically followed their behavior in the Gaussian-noise case to higher values of SNR in those sub-banks in which we found real events.
It is especially remarkable that the sub-bank \texttt{BBH (3,0)} was essentially clean (i.e., without glitches); the five loudest L1 triggers in this sub-bank belonged to GW events that were confirmed using coincident H1 triggers.
Further investigations show that there are dramatic inhomogeneities in the rates at which templates produce triggers that pass our vetoes (i.e., some templates disproportionately trigger on glitches, relative to the bulk). 
Appendix \ref{ap:trigger_hist} presents evidence for this phenomenon.

This is a natural outcome if there is some finite number of `glitch waveforms', in which case only templates that are similar enough to these waveforms produce loud veto-passing triggers (for previous work that reached similar conclusions, see Refs.~\cite{sinegaussian1, sinegaussian2}). 
Guided by this intuition, we identify `glitch-prone' templates using the following empirical procedure:
\begin{enumerate}
    \item Collect all `triggers of interest', defined as L1 triggers with $\rhol^2 > 66$ that pass our vetoes, with the best fit waveform having a chirp-mass $m_{\rm c} > 20 M_{\odot}$, and record the template with the highest value of $\rhol^2$ for each trigger.
    We chose the bound on $\rhol^2$ such that random Gaussian noise would produce (in expectation) only one trigger like this over the entire run: we computed it using the survival function of a chi-squared distribution with five degrees of freedom (amplitude, phase, time, mass, and spin),  $10^2$ independent templates, and 118 days of data. 
    The Gaussian noise hypothesis is unlikely for triggers above this bar, and the remaining explanations are that they are either glitches or genuine signals.
    \item Define as suspected L1 glitches all triggers that pass our vetoes and are not already detected GW events, have $\rhol^2 > 55$ and have available H1 data. We computed this bound in the same way as before, but with one independent template (hence there will be a few Gaussian noise candidates in here, but in practice, glitches dominate this distribution).
    \item For each trigger of interest, count the number of suspected glitches whose templates have a significant match ($\geq 0.9$) with that of the trigger. We use this as an effective measure of the impact of glitches in the associated region of phase-space (note that this implicitly assumes that each template accounts for an equal volume of phase space, which is the prior we adopted in our previous analysis~\cite{Venumadhav:2019lyq}).
\end{enumerate}

We then rank the triggers of interest according to (a) the number of suspected L1 glitches, and then (b) the L1 SNR, $\rhol$. 
We do not assign a higher weight to $\rhol$ in the ranking, since the number of glitches does not steeply decline as a function of SNR (in particular, it does not exhibit the exponential tails characteristic of chi-squared variables).
While this is well motivated, we also tried a few ways of ranking (based on the same criteria as above, but with different parameters), and confirmed that regardless of the choices of numbers (as long as they are high enough to reject the Gaussian contribution), the same set of new triggers joined the set of previously declared events at the top of the list of triggers of interest.

Since the Gaussian noise hypothesis is not viable for the triggers of interest, and we accounted for the effect of glitches in a conservative way (i.e., without over-interpreting high values of $\rhol$), this ranking represents our best degree of belief in a L1 trigger being of astrophysical origin (before considering other detectors, or previous detections). 
Note that our approach differs from previous studies that rank single-detector triggers \cite{2013PhRvD..88b4025C, 2015arXiv150404632C}, since we do not attempt to extrapolate (or indeed model) the distribution of SNRs for glitches, and we account for the extremely inhomogeneous rate at which glitches cause triggers in different parts of our template bank.

Table~\ref{tab:SimilarityRanking} gives the results of this ranking procedure applied to the triggers of interest.

\subsection{Coherent Score from Less-Sensitive Detectors} \label{sec:FishingScore}

The procedure of Section \ref{sec:GlitchSimilarityScore} relies only on the L1 data for a given trigger (note that we use the absence of H1 triggers to build a list of suspected glitches, so the procedure as a whole requires data from Hanford).
For every L1 trigger of interest, we now search for weak counterpart GW signals in other detectors whenever coincident data is available. 

For CBC sources, counterpart signals should agree with the L1 signal in terms of the shape of the waveform, but in general differ in the arrival time, the amplitude normalization, and in the phase constant (this is strictly true only for the dominant (2, 2) harmonic of the GW signal). These are determined by extrinsic parameters, which we denote by the symbol $\Thext$. 

Even when the two LIGO detectors have similar sensitivities, since they are not perfectly anti-aligned, astrophysical sources at special sky locations and with special inclinations can produce disparate SNRs in H1 and L1. 
However, such source configurations are fine-tuned and are thus a priori disfavored. 
For this reason, it is necessary to marginalize over the possible values of the extrinsic parameters. 
For this purpose, we use a conditional coherent score (hereafter coherent score for brevity) $\mathcal{S}$ that we can efficiently compute for each trigger\footnote{Note that this is different from, but analogous to, the coherent score we applied to two-detector coincident triggers in our previous joint search of H1 and L1 triggers, which is an approximation that works in the high SNR limit~\cite{Venumadhav:2019lyq}. \sk{role, in distinguishing between genuine GW signals and coincidental noise, as }}. 

\begin{figure*}[!ht]
    \centering
    \vspace{0.5cm}
    \includegraphics[width=\columnwidth]{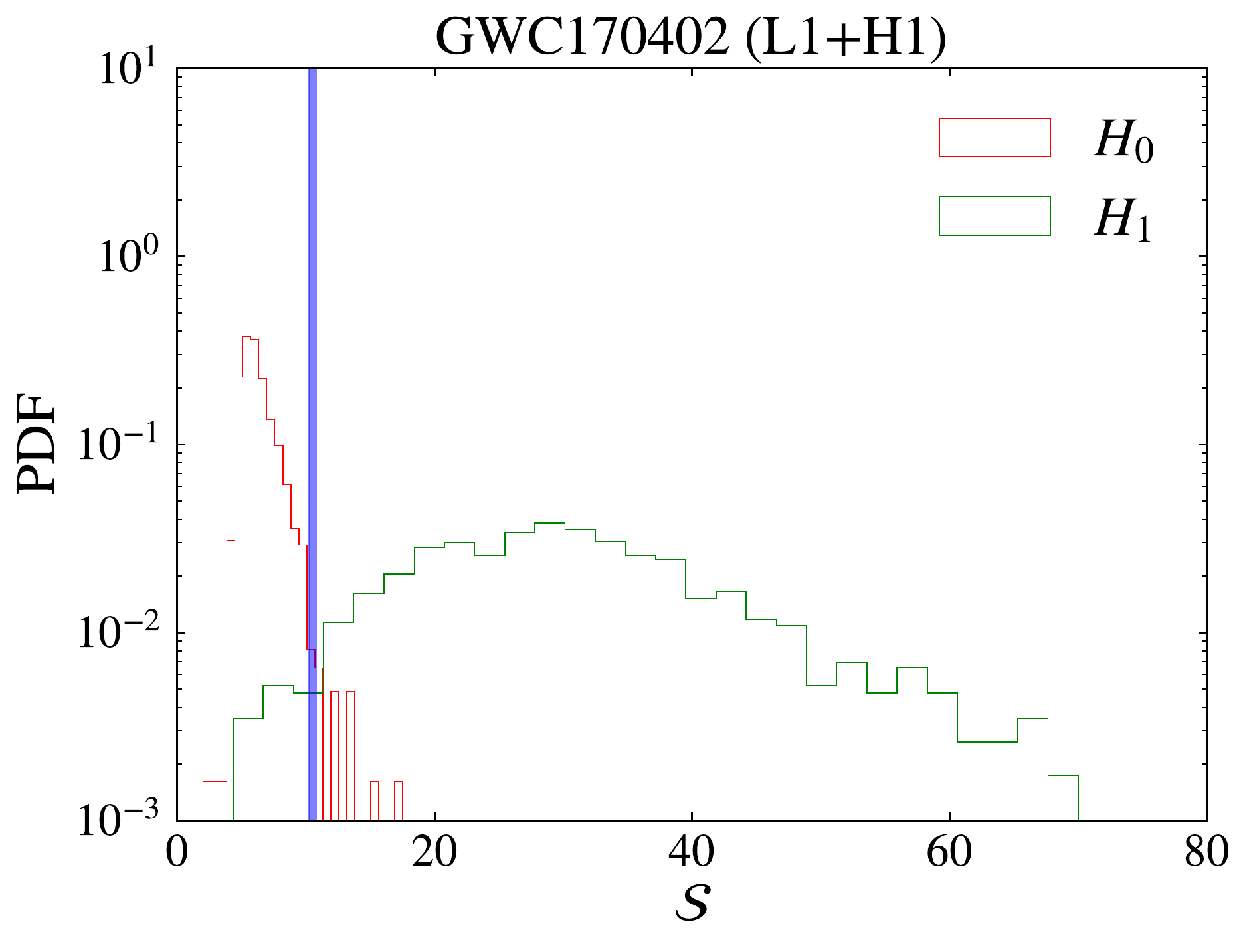}
    \includegraphics[width=\columnwidth]{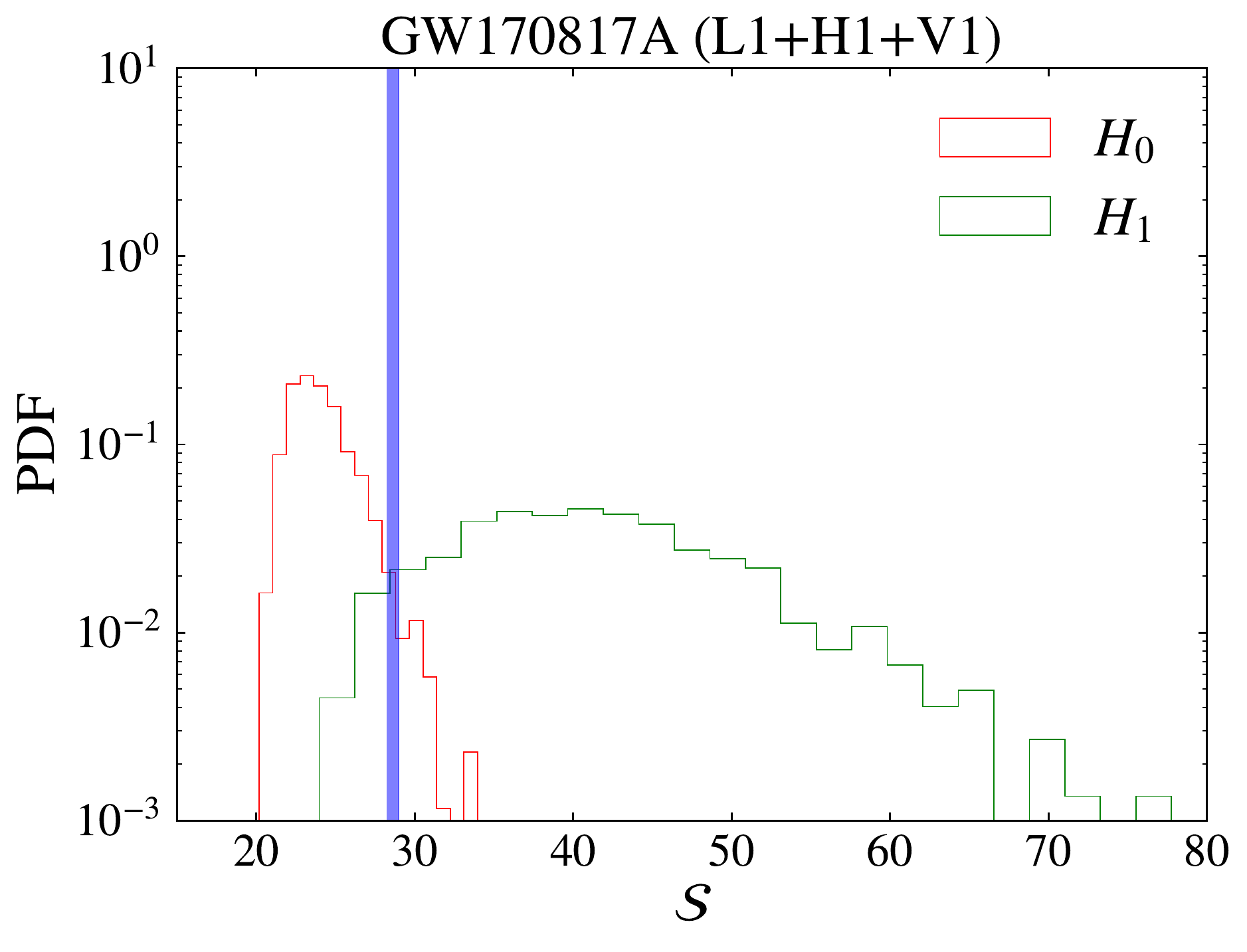}
    \caption{\label{fig:fishing_cand} Distributions of coherent score $\mathcal{S}$ for the top two candidates in Table \ref{tab:SimilarityRanking} using data from the less sensitive detectors (the left and right panels, respectively, show scores for GWC170402 at GPS time 1175205128.565 from H1, and GW170817A at GPS time 1186974148.716 from H1 and V1; see the note in the text about H1 data for GW170817A). 
    The symbols $H_0$ and $H_1$ indicate the noise and astrophysical hypotheses, respectively, under which we derive distributions using 1000 time slides in each case. 
    The vertical shaded region indicates the spread in the values of $\mathcal{S}$ for different choices of the fiducial intrinsic parameters $\Thintr$ consistent with L1 data.}
\end{figure*}

First, we fix the intrinsic CBC parameters, $\Thintr$, (detector-frame masses, spins) to their best-fit values from the L1 data alone. 
The coherent score, $\mathcal{S}$, is the logarithm of the Bayesian evidence for a joint fit to the L1 and H1 data (we can also include V1 data when available), marginalized over all possible combinations of extrinsic parameters $\Thext$:
\be
\label{eq:fishS}
e^{\mathcal{S}} := \int\,\mathcal{D}\Thext\,\Pi(\Thext)\,\mathcal{L}(\bfd |\Thintr, \Thext).
\ee
In the above equation, the symbol $\Pi(\Thext)$ denotes the properly normalized prior for all 7 extrinsic parameters: sky position RA and DEC, polarization angle $\psi$, orbital inclination $\iota$, orbital phase $\varphi$, geocentric arrival time $t_c$, and the source luminosity distance $d_L$. 
The quantity $\mathcal{L}(\bfd |\Thintr, \Thext)$ is the likelihood function, which is given by
\begin{multline}
\ln \mathcal{L}(\bfd |\Thintr, \Thext) = \sum_{i}\,\Big[ \langle d_i | h_i(\Thintr, \Thext) \rangle \\ 
- \frac12\,\langle h_i(\Thintr, \Thext) | h_i(\Thintr, \Thext) \rangle \Big].
\end{multline}
Here, the index $i$ runs over the different detectors, and $d_i$ and $h_i(\Thintr, \Thext)$ are the strain data and the signal in the $i$th detector, respectively. We use $\langle \cdots | \cdots \rangle$ to denote the standard matched filter overlap. 

For the priors $\Pi(\Thext)$, we use a uniform prior distribution on the arrival time $t_c$, and isotropic priors for the source position on the sky and the orbital orientation of the binary. For the luminosity distance $d_L$, we assume a constant volumetric density in Euclidean space within $0 < d_L < 10\,{\rm Gpc}$. In practice, we analytically marginalize over $d_L$ and $\varphi$.

Note that the most rigorous definition of the coherent score (Eq.~\eqref{eq:fishS}) should marginalize over both intrinsic ($\Thintr$) and extrinsic ($\Thext$) parameters, instead of fixing the former to their best-fit values from the L1 data. 
Since the full parameter space is high-dimensional, this significantly increases the the computational cost of evaluating $\mathcal{S}$.\sk{, and we need to determine the distributions $P(\mathcal{S} | H_0)$ and $P(\mathcal{S} | H_1)$ out to their tails separately for each trigger. }
Under the signal hypothesis, the L1 SNR $\rho_{\rm L}$ is high enough to constrain the intrinsic parameters as well as they are in a joint fit to L1 and H1 (and V1 if available) data. 
The values of individual intrinsic parameters (such as masses and spins) are often substantially correlated, but since the degenerate combinations map to nearly the same waveform, and intrinsic parameters are largely uncorrelated with extrinsic ones, it is safe to use the best-fit combination (we computed scores for several different choices of intrinsic parameters consistent with the L1 data, and checked that the answers did not vary significantly).

\sk{That would be the most precise procedure in case the L1 SNR is insufficient so that the values of $\Thintr$ are subject to large uncertainty. }
\sk{Given the large values of the L1 SNR, and the fact that intrinsic and extrinsic parameters are uncorrelated, our choice to fix $\Thintr$ to their best fit values from L1 is practical, and should not lead to biases. }
\sk{In the cases we present, we repeated the procedure for several values of $\Thintr$ taken from their posteriors, and checked that our conclusions were robust. }
\sk{Evaluating $\mathcal{S}$ by full marginalization should be preferred if it can be done more efficiently.}


Given a trigger in the L1 detector, to interpret its associated coherent score $\mathcal{S}$, we need to consider its expected probability density function (PDF) under two competing hypotheses:
\begin{enumerate}
    \item {\it Astrophysical hypothesis ($H_1$):} the L1 trigger is caused by an astrophysical gravitational wave signal, and hence it must have consistent counterpart signals in the other detectors. Under this hypothesis, the coherent score has an expected PDF $P(\mathcal{S} | H_1)$.
    \item {\it Noise hypothesis ($H_0$):} the L1 trigger is caused by noise processes in the detector. In this case, there should not be any correlated counterpart signals in the other detectors. Under this hypothesis, the coherent score has an expected PDF $P(\mathcal{S} | H_0)$.
\end{enumerate}
These distributions inform us about the significance of the event in two ways. 
Firstly, if the L1 single detector trigger is due to an astrophysical event, we expect the coherent score $\mathcal{S}$ to be more consistent with the distribution $P(\mathcal{S} | H_1)$ than with $P(\mathcal{S} | H_0)$. 
Secondly, under the noise hypothesis $H_0$, the probability for $\mathcal{S}$ to be greater than the measured value is analogous to the FAR computed in coincidence analyses~\citep{Venumadhav:2019lyq}.

We determine both the distributions ($P(\mathcal{S} | H_1)$ and $P(\mathcal{S} | H_0)$) locally and independently for each L1 trigger, by empirically sampling from them. 
The local measurement ensures that $P(\mathcal{S} | H_0)$ is representative of each trigger since the noise background in H1 can fluctuate significantly over time. 
More importantly, the relative sensitivities between different detectors vary substantially over the run, which affects the relative strengths of any astrophysical signals within, and hence dramatically changes $P(\mathcal{S} | H_1)$ from one trigger to another\footnote{The same effect is operative in coincidence analyses as well. We accounted for it in Refs.~\cite{Venumadhav:2019lyq, pipeline} using an approximation that is valid in the limit of high SNR in H1.}. 
We can determine the distributions locally and empirically without any extrapolation because we only need to sample tail probabilities $\geq 10^{-3}$.

\sk{This is based on several considerations. First of all, different L1 triggers have different SNRs in L1. Moreover, noise background may fluctuate significantly from one file to another, so it is appropriate to measure a coherent score PDF that reflects the local statistical property of the noise.}

We determine $P(\mathcal{S} | H_0)$ by keeping the L1 strain series fixed, but sliding the strain series in the other detector(s) in time by more than two seconds, which well exceeds the physically allowed time delay relative to L1. 
We then evaluate the coherent score $\mathcal{S}$ as defined previously at this unphysical lag. 
In practice, we restrict the extra time lag to be within a few thousand seconds to obtain a local estimate. 
Our search pipeline also flags ill-behaved segments of data during its preprocessing phase, and masks and in-paints these segments (as well as segments marked by the LVC's quality flags) to avoid contaminating neighboring seconds~\cite{pipeline, Zackay:2019kkv}; we exclude these segments from the time slides.
We repeat this procedure a large number of times and generate samples from $P(\mathcal{S} | H_0)$.

We determine the distribution $P(\mathcal{S} | H_1)$ using a similar procedure as above, with the difference being that we inject counterpart gravitational wave signals into the strain data in the other detector(s) after applying time slides.
We generate injections by fixing the intrinsic CBC parameters $\Thintr$ to their best-fit values, and generating extrinsic parameters $\Thext$ from their posterior distributions (both obtained using only the L1 data). 
We estimate the distribution $P(\mathcal{S} | H_1)$ by repeating the above procedure several times.

Figure \ref{fig:fishing_cand} shows the distributions $P(\mathcal{S} | H_0)$ and $P(\mathcal{S} | H_1)$ for the top two triggers in Table \ref{tab:SimilarityRanking} that are not already confirmed events.
Note that for GW170817A, the spectrogram of the H1 data (i.e., the less-sensitive detector, which was not used to identify the trigger) at the time of the event shows artifacts that are localized to a few bands in the frequency domain. 
Hence, before analyzing the H1 data, we removed frequencies between 68--73 Hz, and 92--96 Hz by applying notch filters (implemented as Bessel filters with critical frequencies at the edges of the quoted frequency intervals). 

\subsection{Determining the False-Alarm Rate}
\label{sec:FAR}

In this section, we describe our procedure for assigning false-alarm rates to single detector triggers. 
We first compute the false alarm probability (FAP) for a trigger of interest, indexed by $i$, given its coherent score $\mathcal{S}_i$ (which is based on the data in H1 and/or V1, conditioned on the loud trigger in L1), as ${\rm FAP}(i) = P(\mathcal{S} > \mathcal{S}_i | H_0) = C(\mathcal{S}_i | H_0)$. 
This is the survival function for the coherent score under the noise hypothesis, $H_0$. 
In order to obtain the false alarm rate, we need to combine this FAP with the occurrence rate of the L1 trigger, and the effective look elsewhere effect.

The triggers of interest were all chosen such that their scores are well above the thresholds for being produced in Gaussian noise.
As we mentioned in Section \ref{sec:GlitchSimilarityScore}, we would like to avoid over-interpreting high values of SNR in L1, since extrapolations of the distribution from lower values of SNR are unreliable. 
Hence, we limit the information from L1 to the rank of the trigger in the list of triggers of interest (this skews to being conservative in interpreting SNR, since the occurrence rates of triggers with a given rank are bounded below by 1 per O2 observing run, and penalizes triggers from regions of parameter space that are affected by glitches).

The rate of triggers being in the first place in the L1 ranking is 1 per O2 (by definition), and hence the probability of the first-place trigger having a ${\rm FAP} < \epsilon$ based on the other detectors is $\epsilon$ per O2. The first trigger on our list has a FAP of 0.015, and hence its false alarm rate is
\begin{align}
     {\rm FAR}^{-1}_{\rm GWC170402} &= \SI{60}{O2} \approx \SI{19}{yr}
\end{align}
Based on this false alarm rate alone, the candidate is well above the threshold significance to be considered interesting~\cite{LIGOScientific:2018mvr}. 
If this trigger also has a high probability of being astrophysical in nature (see more details in Sec.~\ref{sec:PASTRO}), we can add it to the catalog of events, in which case the trigger ranked second becomes the new top candidate (it is standard to remove the background associated with louder events when estimating the significance of fainter triggers, see e.g. Ref.~\cite{PhysRevX.6.041015}). 
The second trigger on the list has a FAP of 0.028, and by a similar argument as above, a FAR of:
\begin{align}
     \rm{FAR}^{-1}_{\rm GW170817A} &= \SI{36}{O2} \approx \SI{11.5}{yr}.
\end{align}
This is also well above the threshold of significance to be considered interesting; we will estimate a value of $p_{\rm astro}$ for this trigger in Section~\ref{sec:PASTRO}.

The triggers further down the list do not have compelling evidence from H1/V1, and hence we terminate the procedure at this point. 
Note that since the two events are at the top of the list, we effectively have no significant corrections due to the look elsewhere effect. 
If events occur further down the list, we would need a more involved procedure that carefully takes the look elsewhere effect into account when estimating their significance. 

\subsection{Determining the Probability that a Trigger is of Astrophysical Origin}
\label{sec:PASTRO}

\begin{figure}[!t]
    \centering
    \includegraphics[width=\linewidth]{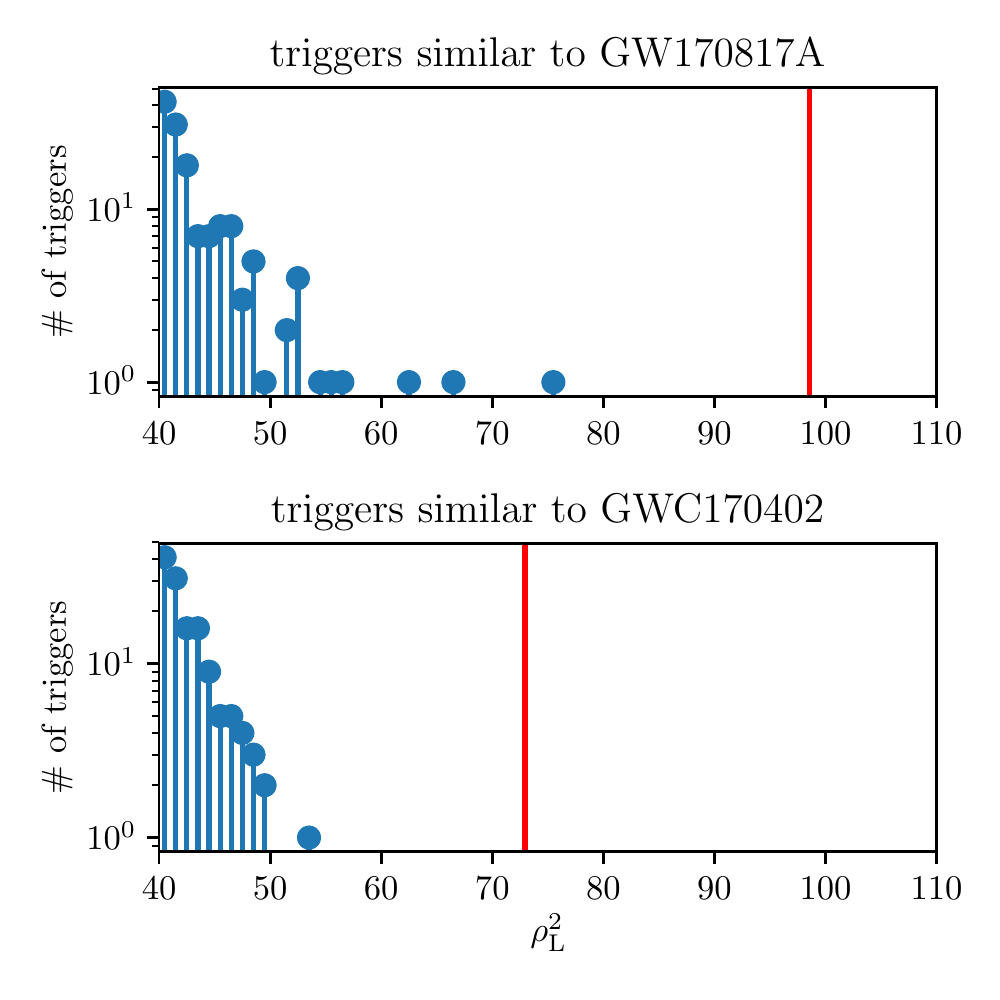}
    \caption{\label{fig:TriggersSimilarToCandidates}     Distributions of L1 $\snr^2$ for triggers for templates that are similar (match $> 0.9$) to the best-fit templates for the two newly found events, that occur at times when the H1 detector is operative.
    Vertical red lines mark the values of $\rhol^2$ for the two events.
    To give context to the amount of phase space that is included in this plot, the upper (lower) panel includes triggers from 28\% (0\%) of bank \texttt{BBH} 4, and 1.8\% (3.6\%) of bank \texttt{BBH} 3.}
\end{figure}

Apart from the FAR, searches in coincident triggers also report a probability of astrophysical origin ($p_{\rm astro}$) for the candidates. 
Given a trigger with a set of properties $\mathcal{T}$, the definition of $p_{\rm astro}$ is:
\begin{align}
    p_{\rm astro} = \frac{R(\mathcal{T}|H_1)}{R(\mathcal{T}|H_0) + R(\mathcal{T}|H_1)} = \frac{\mathcal Q}{1 + \mathcal Q}.
\end{align}
with $\mathcal Q = R(\mathcal{T}|H_1)/R(\mathcal{T}|H_0)$. In order to obtain an estimate of $p_{\rm astro}$ for the candidates in this paper, we would need to empirically measure the distributions of SNR for triggers for similar templates in Livingston, and extrapolate them to higher values (the candidates we are discussing are the loudest triggers in their distributions, see Fig.~\ref{fig:TriggersSimilarToCandidates}).
As we mentioned in the introduction, we do not have a physical model for glitches, due to which the results of this extrapolation are subject to significant uncertainties. 

In a desire not to over-interpret the high values of $\rhol^2$, we conservatively restrict the information from L1 to the fact that $\rhol^2 > 66$, and that the corresponding templates are in the `clean', or `glitch-free' region of parameter space. Note that all previously discovered black hole mergers are comfortably inside the `clean' region of parameter space.

We then proceed with the calculation of the rate ratio,
\begin{equation}
    \mathcal{Q} = \frac{R(\mathcal{T} | H_1)}{R(\mathcal{T} | H_0)} = \frac{R(\rhol^2 > 66, {\rm  clean} | H_1)}{R(\rhol^2 > 66, {\rm  clean} | H_0)}\,\frac{P(\mathcal{S}|H_1)}{P(\mathcal{S}|H_0)},
\end{equation}
where the first and second terms in both the numerator and denominator give the information coming from the stronger detector and that from the weaker detectors. 
The rate
\ba
R(\rhol^2 > 66, {\rm  clean}|H_0)
\ea
effectively represents the rate of triggers that are above the trigger in question (including itself) in the Livingston based ranking. 
Since these two triggers are at the top of the list, we can set this number to one per O2. 
We caution the reader that this estimate is very uncertain, because we cannot run the experiment of making such a rank on fake data many times (as we cannot simulate real single-detector data). 
We should consider this number as being subject to order unity uncertainty, since, in principle, there could have been a glitch above our events in the rank.
\begin{multline}
    R(\rhol^2 > 66,{\rm  clean}|H_1) \\ 
    = \mathcal{R_{\rm O2}} \times \frac{P(\rhol^2 > 66)}{P(\rhoh^2 + \rhol^2 > \rho^2_{\rm threshold}, \rho_{\rm H, L}^2>16 )},
    \label{eq:rh1}
\end{multline}
where $\mathcal{R_{\rm O2}} = 13$ is the total reported rate for events in banks \texttt{BBH 3} and \texttt{BBH 4}, as reported in \cite{Venumadhav:2019lyq}, and we account for the different volumes that the two analyses are sensitive to. 
We use the combined rate because the analysis in this paper covers the union of these two banks (as we mentioned at the beginning of Sec.~\ref{sec:Methodology}).
The second term on the right-hand side of Eq.~\eqref{eq:rh1} is the ratio of sensitive volumes for the coincidence and single-detector analyses, and depends on the relative sensitivities of the detectors in the network (we only consider H1 and L1 when estimating significance). 
This volume ratio evaluates to 0.33 (0.5) when the detectors are equally sensitive (L1 is $40\%$ more sensitive than H1) as is relevant to the case of GWC170402 (GW170817A).

Substituting all the above factors, the rate ratios for the two events evaluate to:
\begin{equation}
    \begin{split}
        \mathcal{Q}_{\rm GWC170402} &\approx 2.15 \\
        \mathcal{Q}_{\rm GW170817A} &\approx 6.37,
    \end{split}
\end{equation}
which implies that their probabilities of being of astrophysical origin are:
\begin{equation}
    \begin{split}
        p_{\rm astro}({\rm GWC170402}) &= 0.68 \\
        p_{\rm astro}({\rm GW170817A}) &= 0.86.
    \end{split}
\end{equation}
Note that this does not include any information from the fact that these triggers are outliers in their respective, locally estimated background distributions. 
If we had a credible way of accounting for this fact, it would only increase the inferred values of $p_{\rm astro}$. 
Figure \ref{fig:TriggersSimilarToCandidates} shows the local background distributions for triggers for templates similar to those for the two events. 

\subsection{Validation using GW170818}
\label{sec:170818}

\begin{figure}[!t]
    \centering
    \vspace{0.5cm}
    \includegraphics[scale=0.45]{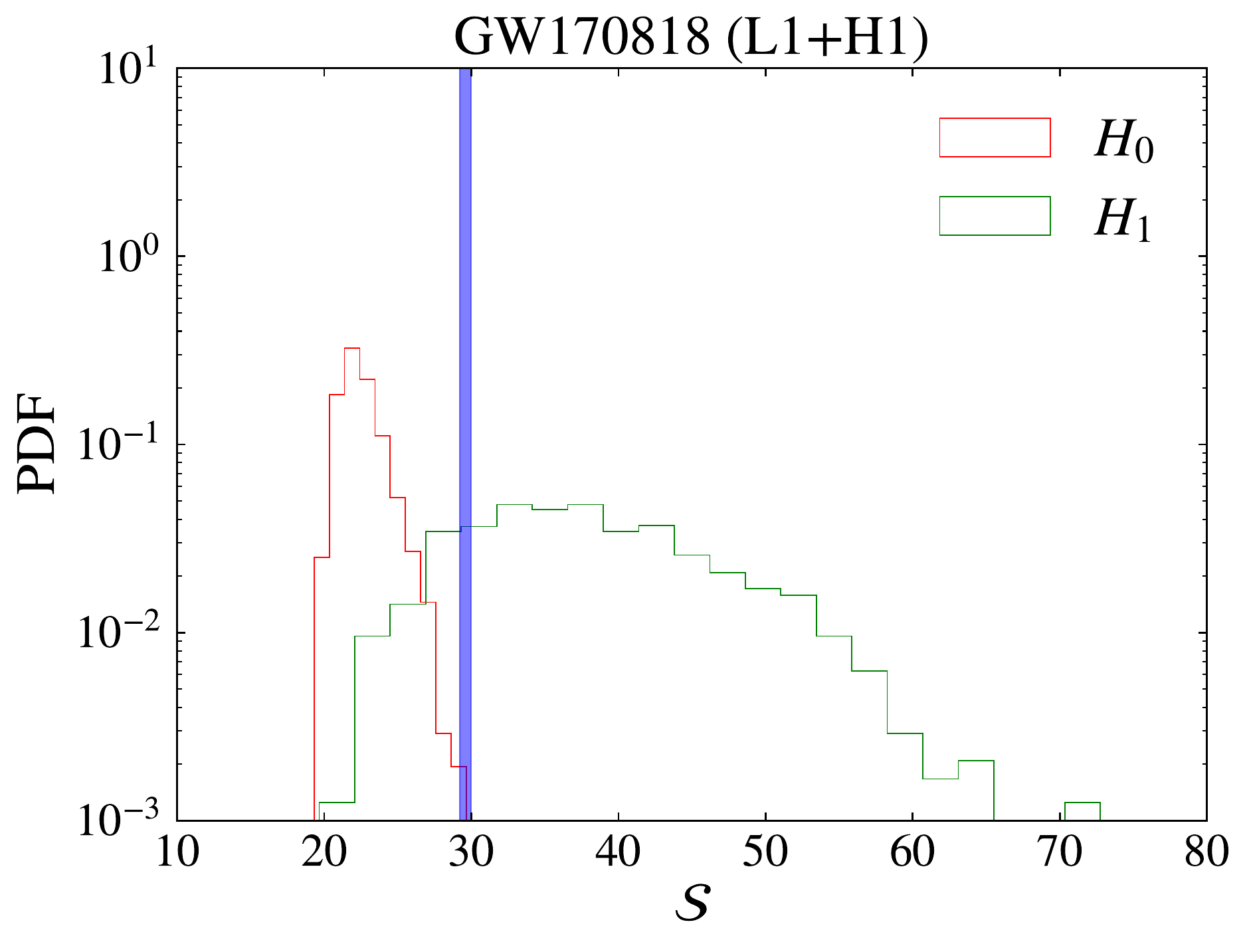}
    \caption{\label{fig:gw170818} Demonstration of our coherent score $\mathcal{S}$ with the LVC event GW170818 using data from only the two LIGO detectors. The notation, and the number of time slides used, are identical to those of \reffig{fishing_cand}.}
\end{figure}

\begin{figure*}[!ht]
    \centering
    \includegraphics[width=\linewidth]{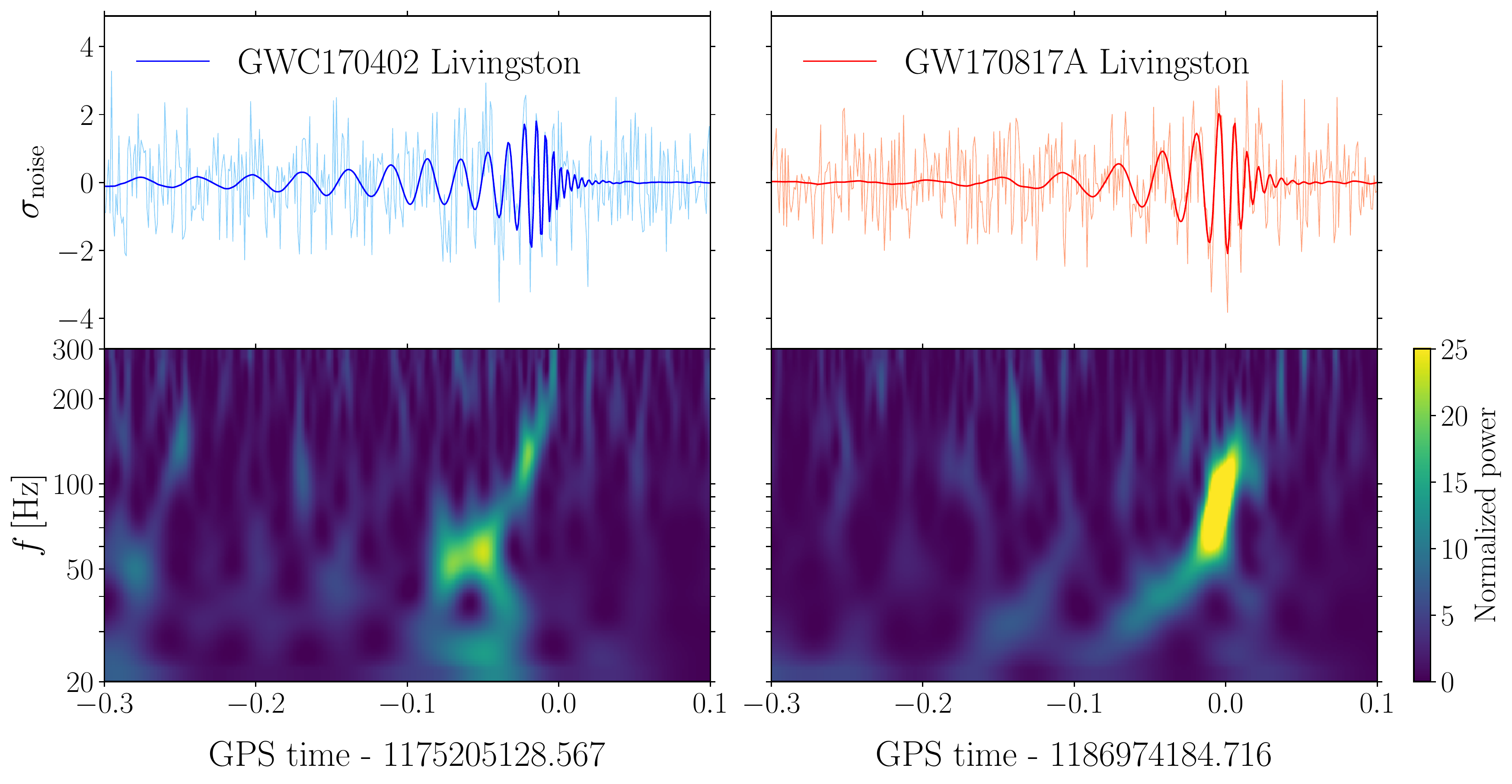}
    \caption{\label{fig:LST_two_cand_wht_spec} Two BBH candidates GWC170402 and GW170817A initially selected as significant L1 single detector triggers. Upper panels show the whitened strain series around the trigger times (light colored curves), with the network maximum likelihood \texttt{IMRPhenomD} waveforms overplotted (dark colored curves). The corresponding spectrograms are shown in the lower panels.}
    \label{fig:LST_two_cand_wht_spec}
\end{figure*}

We also validate the above procedure by applying it to GW170818, a BBH event previously reported by the LVC~\citep{LIGOScientific:2018mvr}. It was marked as an L1 single detector trigger by the \texttt{PyCBC} pipeline but was not considered for coincidence analysis because its SNRs at H1 and V1 were below the threshold for collection. It was initially identified by the \texttt{GstLAL} pipeline as a L1--V1 double detector trigger until a H1 counterpart signal was later confirmed in the offline search. It was also confirmed by a refined analysis with the \texttt{PyCBC} pipeline \cite{2ogc}. 

GW170818 is an ideal example to demonstrate how the astrophysical nature of a single detector trigger can be validated by evaluating the coherent score, as it has a high SNR at L1 ($\sim 10$) but very low SNRs at H1 and V1 (both $\sim 4$). 
Figure \ref{fig:gw170818} shows the coherent score $\mathcal{S}$ for this event using the data in L1 and H1 alone, and its distributions under the noise and astrophysical hypotheses. 
The high value for the coherent score, $\mathcal{S}=29.7$, measured from the consistency of the data in the two LIGO detectors, is an obvious outlier relative to $P(\mathcal{S}|H_0)$, with none of our 1000 Montecarlo realizations yielding a higher value for the score. 
In contrast, it is fully consistent with typical values drawn from $P(\mathcal{S}|H_1)$. At $\mathcal{S}=29.7$, the probability density for the coherent score under the astrophysical hypothesis is more than 30 times higher than that under the noise hypothesis. 

In Table \ref{tab:SimilarityRanking}, this is the trigger with the highest L1 SNR among those that have not been validated in the two-detector joint analysis, and it has no similar glitches. 
Considering these facts, we are able to assign an inverse FAR better than \SI{1000}{O2} for GW170818 purely from the coherence of the recovered signals in L1 and H1. 
Hence, we are able to confirm its astrophysical origin even without confirmation from the Virgo detector. 

\sk{We we apply this procedure to a long list of single detector triggers, vast majority of which are expected to be glitches as reflected by their L1 based ranking.
In Fig. \ref{fig:FishingSurvivalFunction} we present the survival function distribution of the triggers we ran.}

\section{Summary of Results}
\label{sec:results}

Table \ref{tab:SimilarityRanking} summarizes our results on O2. 
Among the eight highest (L1-based) ranking events, six were previously detected BBH events (that were detected using Hanford (H1) and Virgo (V1) data), which gives us some confidence in the ability of the ranking statistic to identify interesting triggers. 

For the other two triggers, we detect faint counterpart signals in Hanford with a locally measured inverse false-alarm rate of ${\rm FAR}^{-1} > \SI{36}{O2} \approx \SI{11.5}{yr}$. 
Note that for GW170817A, the H1 data needs to be cleaned of artifacts in the frequency domain by notching out frequencies 68--73 Hz, and 92--96 Hz.

We tested triggers further down on the list, and do not find any significant supporting evidence from the Hanford detector for any of them. 
We additionally verified that the distribution of the computed scores (quantifying multi-detector coherence) for these fainter triggers is consistent with the predicted one with no signal (and hence consistent with pure background events in the Livingston detector).

We name the two new events GWC170402 and GW170817A according to the date of occurrence. 
Figure \ref{fig:LST_two_cand_wht_spec} shows the spectrograms of the strain data in the Livingston detector around the two events.

\section{Astrophysical Implications of GW170817A}
\label{sec:astro}

\begin{figure}[!t]
     \centering
     \includegraphics[width=\columnwidth]{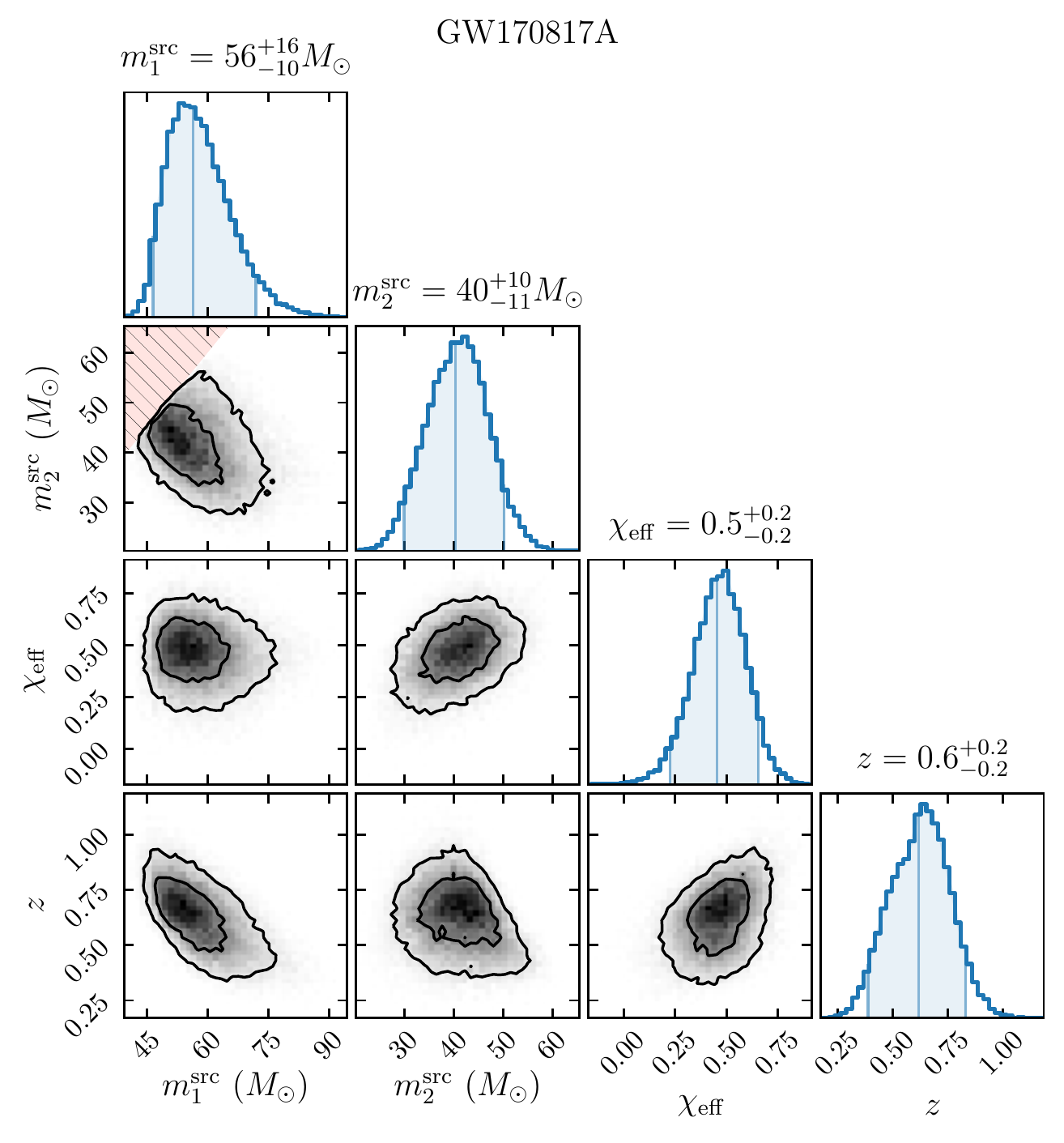}
     \caption{Marginalized posteriors for GW170817A. Two-dimensional contours enclose 50\% and 90\% of the distribution. In the one-dimensional posteriors, vertical lines mark the 0.05, 0.5 and 0.95 quantiles. We compute likelihoods using the \texttt{IMRPhenomD}~\cite{Khan2016} waveform model, and adopt a prior that is uniform in detector-frame $m_1$, $m_2$, $\chi_{\rm eff}$ and luminosity volume.}
     \label{fig:Posteriors_m1m2}
\end{figure}

We perform parameter estimation for GW170817A by using relative binning \cite{Zackay2018} to compute the likelihood, and the PyMultiNest code \cite{PyMultiNest} to generate samples from the posterior. 
As noted earlier, we had to apply notch filters to the H1 data to determine the significance of this event; we apply the same filters before performing parameter estimation as well.
Figure \ref{fig:Posteriors_m1m2} presents posteriors for the source-frame masses, effective spin, and the redshift, marginalized over other parameters.

The inferred source frame total mass of GW170817A is $M^{\rm src}_{\rm tot} = 98^{+17}_{-11}\,M_\odot$, while its inferred effective spin is $\chi_{\rm eff} = 0.5^{+0.2}_{-0.2}$. 
It is interesting to consider the individual component masses: the primary black hole has a source-frame mass of $m_1^{\rm src} = 56_{-10}^{+16} \, M_\odot$ (the limits indicate $95 \%$ confidence intervals), which would put it at the heaviest end of the merging black holes discovered so far (see Ref.~\cite{LIGOScientific:2018mvr}). 

The presence of such massive individual black holes is potentially informative about the formation of stellar mass black holes from massive progenitor stars at the end of their lives. 
Models of stellar evolution predict that extremely massive stars with Helium core masses in the range of $\sim 50$--130\,$M_\odot$ do not produce BH remnants with these masses, since they either explode as Pair Instability Supernovae and leave no remnants, or shed substantial mass via the Pulsational Pair Instability and leave lower-mass remnants~\cite{2017ApJ...836..244W}. 
However, the location of this mass gap is subject to large uncertainties, since it depends on the phenomenology of mass loss~\cite{2017MNRAS.470.4739S}. 
Extremely metal-poor stars may collapse into BH remnants even as massive as 70--80 $M_\odot$ \cite{2017ApJ...836..244W, 2018ApJS..237...13L}. 
It has also been suggested that dense stellar systems can harbor massive BBHs in which one (or both) of the components is a product of a prior merger, which would produce BHs above any mass gap~\cite{2019PhRvD.100d3027R}.

Previous work has used the LVC detections to jointly infer the properties of the population of BBHs in the Universe~\cite{Fishbach2017,Talbot2018,Wysocki2019,Roulet2019,lvcpopulation}.
Partially motivated by the above astrophysical considerations, some of the models considered have an upper cutoff to the mass of the merging BHs: the inferred cutoff mass is at $\sim 42$--44\,$M_\odot$ (effectively the lowest end of the posterior of the most massive system, GW170729), with a tail toward higher values. 
If GW170817A were incorporated into such an analysis, the inferred cutoff would be at higher values of the mass and the constraint would be strengthened compared to using just GW170729 alone. 

\begin{figure*}
    \centering
    \includegraphics[width=\linewidth]{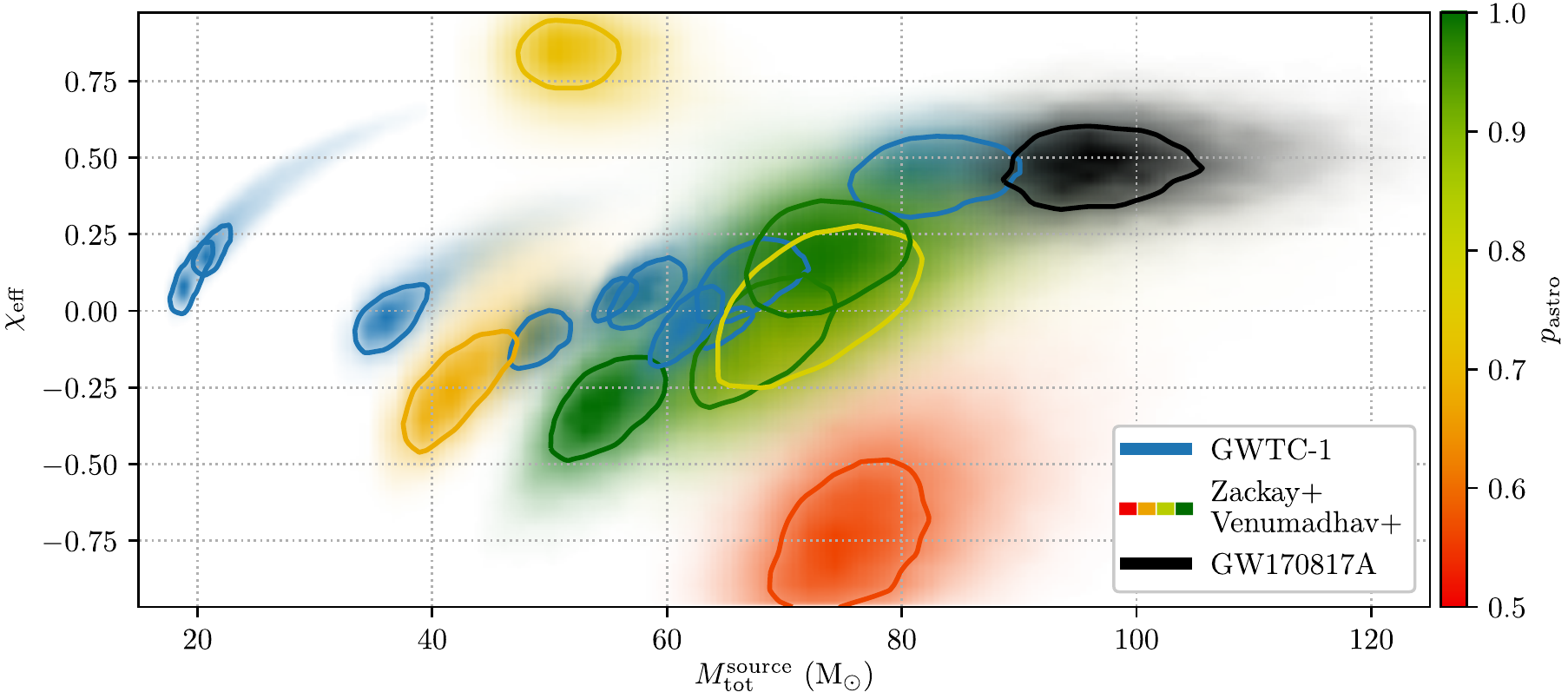}
    \caption{Binary black holes events reported from O1 and O2, in the plane of source-frame total mass vs. effective spin. In blue are shown the 10 BBH events reported in GWTC-1~\cite{LIGOScientific:2018mvr}, all of them are certainly astrophysical in origin ($p_{\rm astro} = 1$). Color coded by $p_{\rm astro}$ are shown 7 additional events with $p_{\rm astro} > 0.5$ that our previous searches found~\cite{GW151216, Venumadhav:2019lyq}. In black we show GW170817A. Displayed are $1\sigma$ probability contours, i.e. enclosing $1 - e^{-1/2} \approx 0.39$ of the probability distribution.}
    \label{fig:all_events}
\end{figure*}

Finally, Figure \ref{fig:all_events} places the event in the context of all the other events detected thus far in terms of the total source-frame mass and effective spin parameter, $\chi_{\rm eff}$. 
Interestingly, GW170817A follows the emerging trend of the total source-frame mass and the effective spin being correlated at the heavier end of the detected population of events. 
We need a proper accounting of the selection effects of the search pipelines to assess the astrophysical relevance of any putative correlations. 

\section{Conclusions}
\label{sec:conclusions}

In this paper, we presented and applied a new method to assess the false-alarm rates of loud single detector triggers that have weak counterpart signals in other detectors. 
This method is motivated by the fact that there is significant sensitive volume in this regime when the detectors that make up the network have disparate sensitivities, such as was the case during the O2 run, and that analyses of coincident triggers do not cover this volume due to cuts on SNR. 
We also note that in the ongoing O3 run, the Livingston detector continues to be substantially more sensitive than the others, and hence we expect that the method we present will be important even in the future.
It could be especially interesting to apply this technique to several binary neutron star candidates identified thus far by the LVC during O3 that have only Livingston--Virgo co-detection\footnote{\url{https://gracedb.ligo.org/superevents/public/O3/}}.

When we apply our method to data from the O2 observing run, we detect two additional significant events. 
One of the events, GW170817A is the merger of a pair of very massive black holes; its estimated parameters suggest that it could be the most massive merger reported so far. It has been theoretically suggested that stellar mass BHs are subject to a mass cutoff at $\sim 40$--$50\,M_\odot$ due to the physics of pulsational pair instability supernovae and pair instability supernovae of the progenitor star~\cite{1964ApJS....9..201F, barkat1967dynamics}. 
GW170817A should be very valuable in constraining the existence and the exact location of such a mass cutoff~\cite{Belczynski:2016jno, Woosley:2016hmi, Spera:2017fyx, Marchant:2018kun, Stevenson:2019rcw}. 

The other event, GWC170402, if genuine, is perhaps the most interesting as it shows hints of additional physics not included in the waveform models we have used in this paper. 
We will present a detailed analysis of this event in a companion paper. 

It is also interesting that the distribution of masses and spins of the events detected so far indicates a correlation between the total source-frame mass and the effective spin parameter, $\chi_{\rm eff}$. 
We need more detections and a careful population analysis to confirm the astrophysical nature of this correlation; if real, it may shed light on the binary stellar evolution of massive stars.


\section*{Acknowledgments}

We thank the LIGO--Virgo Collaboration for making the O1 and O2 data publicly accessible and easily usable.

This research has made use of data, software and/or web tools obtained from the Gravitational Wave Open Science Center (\url{https://www.gw-openscience.org}), a service of LIGO Laboratory, the LIGO Scientific Collaboration and the Virgo Collaboration. LIGO is funded by the U.S. National Science Foundation. Virgo is funded by the French Centre National de Recherche Scientifique (CNRS), the Italian Istituto Nazionale della Fisica Nucleare (INFN) and the Dutch Nikhef, with contributions by Polish and Hungarian institutes.

BZ acknowledges the support of Frank and Peggy Taplin Membership Fund. LD and TV are supported by John Bahcall Fellowships at the Institute for Advanced Study. MZ is supported by NSF grants AST-1409709,  PHY-1521097 and  PHY-1820775 the Canadian Institute for Advanced Research (CIFAR) Program on Gravity and the Extreme Universe and the Simons Foundation Modern Inflationary Cosmology initiative.


\appendix

\section{Inhomogeneous distribution of glitches}
\label{ap:trigger_hist}

\begin{figure*}[!ht]
\begin{minipage}{0.45\textwidth}
\centering
    \begin{tabular}{|c|c|c|c|c|}
    \hline
        \rule{0pt}{3ex} Bank ID & \# $\rhol^2>45$ & \# $\rhol^2>55$    & \# $\rhol^2>65$  \\[1ex]
        \hline
        (2,0) & 10178 & 172 & 1\\
        (2,1) & 1558 & 50 & 7\\
        (2,2) & 734 & 226 & 102\\
        (3,0) & 337 & 18 & 7\footnote{Six of the seven triggers in bank (3,0) are previously declared gravitational wave signals. The seventh is declared in this paper}\\
        (3,1) & 157 & 11 & 3 \\
        (3,2) & 41 & 8 & 4 \\
        (4,0) & 37 & 3 & 1\footnote{This trigger is GW170823} \\
        (4,1) & 14 & 1 & 0 \\
        (4,2) & 9 & 3 & 2 \\
        (4,3) & 32 & 11 & 4 \\
        (4,4) & 215 & 77 & 31\\
        \hline
    \end{tabular}
\end{minipage}
\begin{minipage}{0.45\textwidth}
\centering
    \includegraphics[width=80mm]{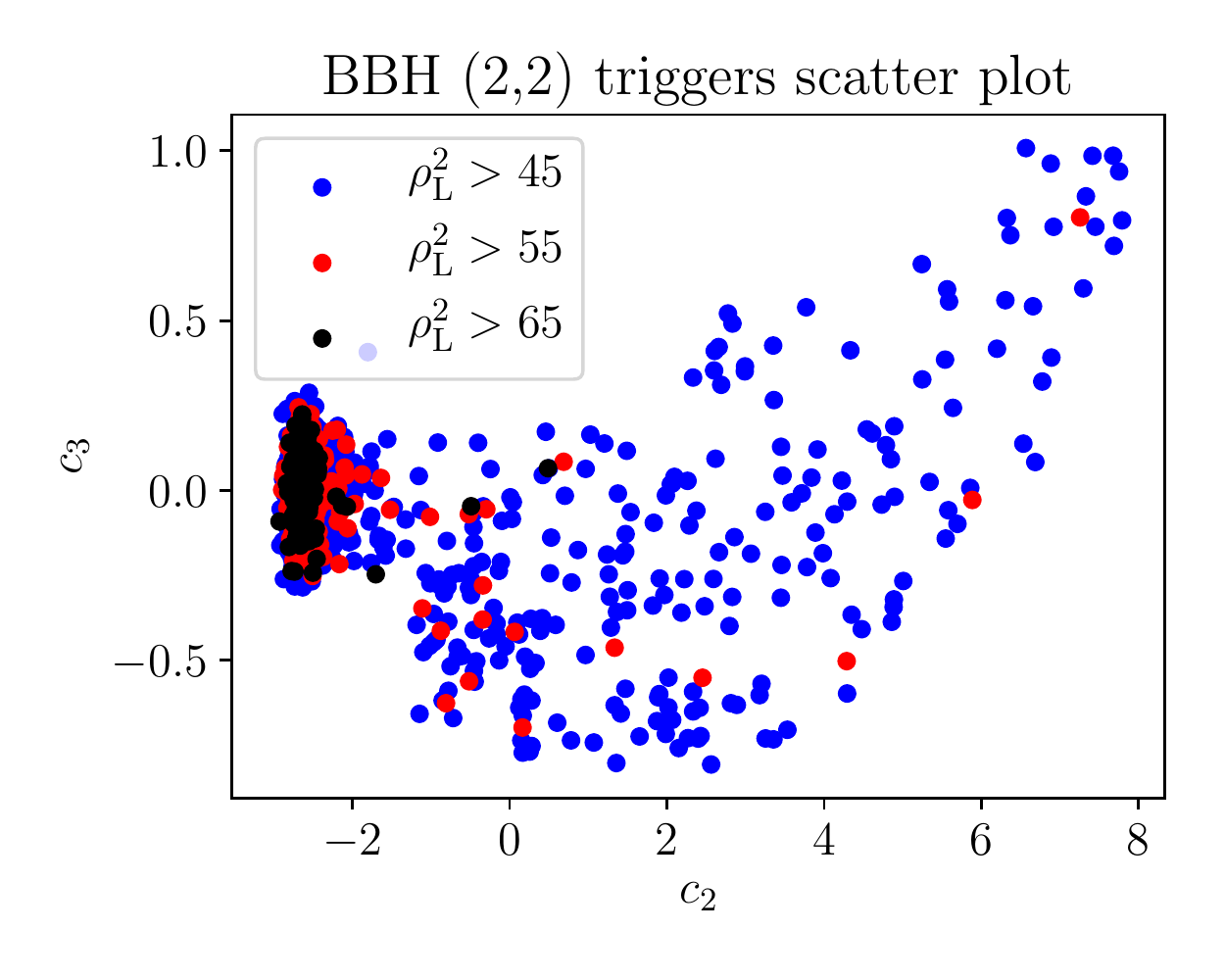}
\end{minipage}
\caption{\label{triggers-vs-subbank} The table in the left-hand panel shows the number of veto-passing L1 triggers in each sub-bank above a few threshold values of the SNR. The non-uniform numbers of triggers with $\rhol^2 > 65$ shows that glitches are localized within certain sub-banks. The plot in the right-hand panel shows the coefficients labeling the templates for triggers above the thresholds for bank \texttt{BBH} (2, 2). Note that glitches are localized within a small region of parameter space.}
\end{figure*}

In Section \ref{sec:GlitchSimilarityScore}, we provided a ranking of single-detector (L1) triggers. 
The ranking relied on the empirical observation that after we applied signal-quality vetoes to the triggers that the matched filtering procedure returned, the remaining glitches were confined to certain `glitch-prone' regions within the set of templates that we used. 

The table and associated figure in Fig.~\ref{triggers-vs-subbank} present evidence of this effect. 
The table in the left-hand panel shows the numbers of veto-passing triggers above three threshold values of $\snr^2$ (45, 55, and 65) for the heavier banks and sub-banks that we used in our search.
Note that the templates in bank \texttt{BBH 2} and its subbanks cover signals with chirp-masses $m_{\rm c} \in (12, 20) \, M_\odot$, while the search in this paper covers signals with $m_{\rm c} > 20 \, M_\odot$. 
We include this bank because its sub-bank \texttt{BBH (2,2)} shows the most dramatic example of the phenomenon of localized glitches.

At low values of the SNR (the first column in the table), the numbers are controlled by Gaussian noise, and hence the disparity in numbers largely reflects the different numbers of templates in the various banks/subbanks (except \texttt{BBH} (4, 3) and (4, 4), which show signs of glitches even at $\snr^2 = 45$). 
At larger values of the SNR, the distributions are dominated by glitches, and we can see that the effects are localized to within a few subbanks. 
Even within subbanks, there are a few glitch-prone templates that dominate the tail of the distribution. 
The figure in the right-hand panel of Fig.~\ref{triggers-vs-subbank} is a scatter plot of the first two coefficients that index our template bank for the triggers in \texttt{BBH} (2, 2). 
We see that almost all the glitches are localized to a small region within the bank (as shown by the red and black markers, which are the triggers with $\rhol^2 > 55$ and $65$, respectively.

\section{A Spurious Candidate}
\label{sec:glitch_specgram}

\begin{figure*}
    \centering
    \includegraphics[width=160mm]{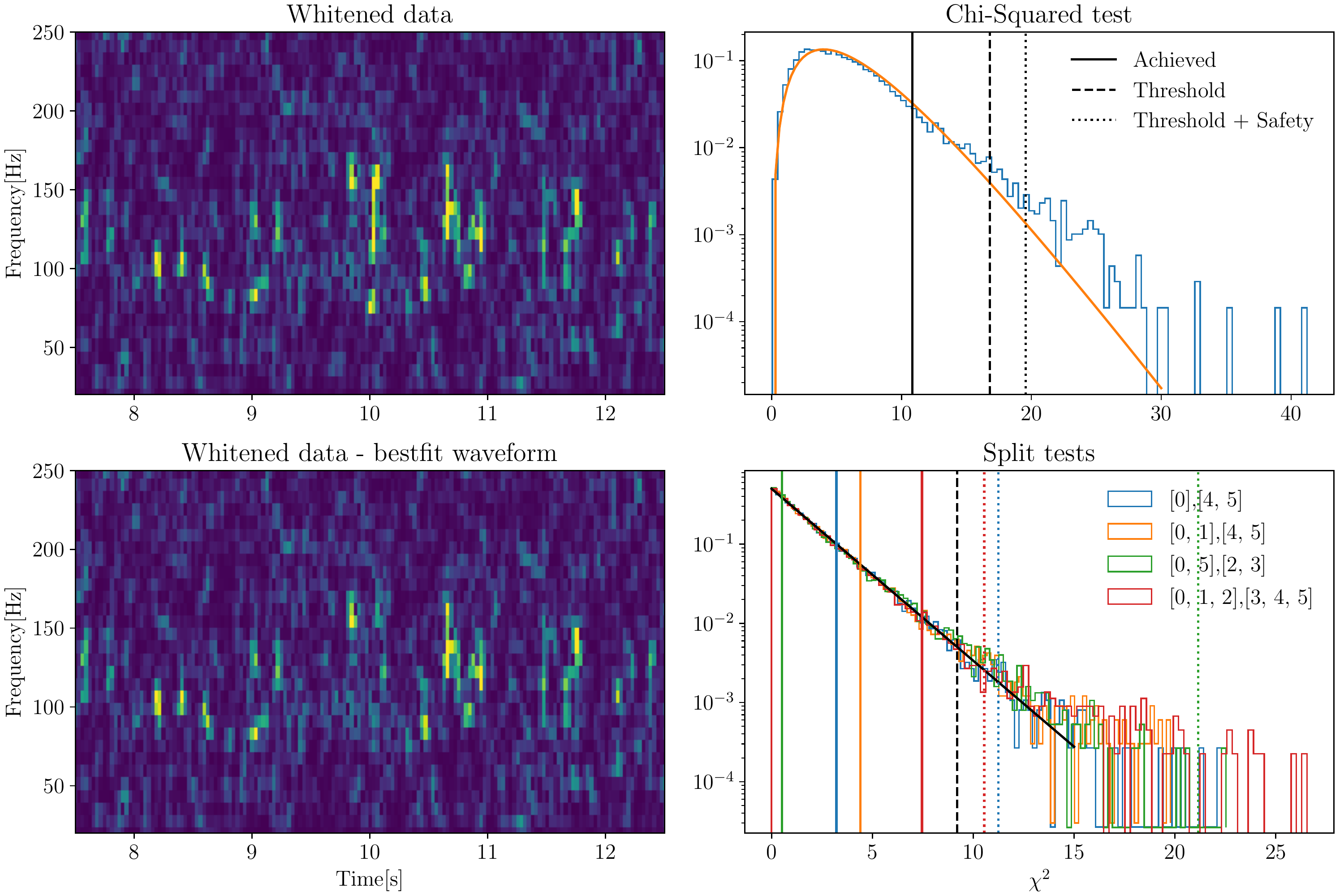}
    \caption{Details of the glitch appearing in the last row of Table \ref{tab:SimilarityRanking}.\textit{ Top left:} Spectrogram of the data around the trigger, the merger time is at 10 seconds. It is visually obvious that the event is a glitch, due to the activity in the few seconds around it.\textit{ Bottom left:} Spectrogram of the data with the best fit template removed. The chirp itself is of sufficiently high quality that the subtracted region cannot be rejected by itself, without the context of the surrounding data.
    \textit{ Top right:} Results of a chi-squared veto based on the matched-filter overlaps of subsets of the whitened waveform.
    \textit{ Bottom right:} Results of split-tests based on the matched-filter overlaps of a few combinations of subsets of the whitened waveform. Each color corresponds to a different test. See text for details.
    The histograms show off-event distributions of all the test-statistics we use to veto the trigger, and the solid lines show the their predicted distributions. 
    The solid vertical lines are the values of the test statistic that are achieved by the event, and the dotted vertical lines show the thresholds applied on the test statistics (triggers to the left are retained to avoid rejecting real events).}
    \label{fig:GlitchSpecGram}
\end{figure*}

The ranking in Sec.~\ref{sec:GlitchSimilarityScore} marked the L1 triggers for all previous loud events, and produced a short list of remaining single-detector candidates. 
In Tab.~\ref{tab:SimilarityRanking}, we noted that the ranking procedure produced a candidate that had clear artifacts in its spectrogram. 
Figure \ref{fig:GlitchSpecGram} presents the spectrogram for the L1 data around the time of this candidate. 
Our automated pipeline relies on a series of tests to reject glitches, and Fig.~\ref{fig:GlitchSpecGram} includes the results of these tests. 
The lower-left panel shows the test performed to check whether the signal-subtracted data shows excess power -- since the non-stationarity persists on longer timescales, and the test checks against a local average, this candidate was not rejected.
The right-hand panels show the results of the vetoes that test consistency between the matched-filtering scores of sub-chunks of the best-fit whitened waveform; the results for this candidate are within the thresholds that we impose based on our requirements not to veto real signals. 
We divide the best-fit whitened waveform into six chunks with equal values of SNR, and test for the consistency of the matched-filtering overlaps of these chunks. 
The top-right panel shows the results of a chi-squared-like test that tests consistency between all six chunks \cite{PhysRevD.71.062001}, and the bottom-right panel shows the results of split tests that test consistency between certain combinations of the chunks (the designation $[a,\dots],[b,\dots]$ denotes tests in which we compare the set of overlaps $z_a, \dots$ to the set $z_b, \dots$). 
More details of this procedure will be provided in a future paper~\cite{vetopaper}.

While these tests help reduce the effects of glitches, they are not perfect since they were informed by our previous experiences looking at small subsets of the data. 
In principle, we can design a test with criteria such that we can better reject this candidate and any other candidates like it, and add it to the battery of tests we have. 
We choose not to do so, because we do not have several examples of this glitch to measure the selectivity of any tests, and more importantly, it would make our analysis less blind. 
In this particular case, even if the tests do not reject the candidate, there is enough obvious non-stationary behavior that we can visually reject the possibility that it is of astrophysical origin. 

\bibliographystyle{apsrev4-1-etal}
\bibliography{gw}

\end{document}